\documentclass[pdftex, preprint, showpacs,preprintnumbers,amsmath,amssymb, showkeys, pre]{revtex4}

\usepackage[]{graphicx}
\usepackage{dcolumn}
\usepackage{array}
\usepackage{stmaryrd}
\usepackage{color}
\begin{document}

\definecolor{orange}{rgb}{1, 0.5, 0}

\title[Vortex ring refraction]{Vortex ring refraction at large Froude numbers}

\author{Kerry Kuehn}
\email{kerry.kuehn@wlc.edu}
\homepage{http://web.me.com/kerry_kuehn/Physics}
\author{Matthew Moeller}
\author{Michael Schulz}
\author{Daniel Sanfelippo}
\affiliation{Department of Physical Sciences, Wisconsin Lutheran College, Milwaukee, WI 53226}

\date{\today}

\begin{abstract}
We have experimentally studied the impact of an initially planar axisymmetric vortex ring, incident at an oblique angle, upon a gravity-induced interface separating two fluids of differing densities.  After impact, the vortex ring was found to exhibit a variety of subsequent trajectories, which we organize according to both the incidence angle, $\theta_i$, and the interface strength, defined as the ratio of the Atwood and Froude numbers, $A/F$. For grazing incidence angles ($\theta_i \gtrsim 70$ deg.) vortices either penetrate or reflect from the interface, depending on whether the interface is weak or strong.  In some cases, reflected vortices execute damped oscillations before finally disintegrating.   For smaller incidence angles ($\theta_i \lesssim 70$ deg.) vortices penetrate the interface.  When there is a strong interface, these vortices are observed to curve back up toward the interface.  When there is a weak interface, these vortices are observed to refract downward, away from the interface.  The critical interface strength below which vortex ring refraction is observed is given by  $\log_{10}{(A/F)}= -2.38 \pm 0.05$.  
\end{abstract}

\pacs{47.32.C-, 42.15.-i, 42.50.Ct, 47.55.dr}


\maketitle

\section{Introduction}
\label{sec:intro}

Due to their consistent appearance in many hydro- and aero-dynamic systems, vortex rings have been studied in the laboratory for decades~\cite[][]{Maxworthy:1972p15}. 
Some studies have focused on the interaction of a vortex ring with either a solid surface~\cite[][]{Lim:1989p1676} or a free surface~\cite[][]{Bernal:1989p1810, Sarpkaya:1996p1875}. Others have focused on the trajectory of a vortex ring when propagating through fluid with a gradual vertical density gradient~\cite[][]{Honji:1976p10, Scase:2006p302}.  

Comparatively few studies have been performed on the interaction of a vortex ring with an interface separating two fluids of different densitites.   Among these, \citet*[][]{Linden:1973p11} studied the normal incidence of a vortex ring on a fluid density interface.  He found that the penetration depth of the ring into the lower, more dense fluid, depended upon the Froude number of the vortex ring.  \citet*[][]{Dahm:1989p8} performed a comprehensive study of the normal incidence of a vortex ring on a fluid density interface.  They found that for thin interfaces, the interaction of the vortex ring with the interface was governed by the Froude number, $F$, and the Atwood number, $A$. Furthermore, they found that in the Boussinesq limit ($A \! \shortrightarrow \! 0$) the interaction could be characterized by the ratio $A/F$, which represents a dimensionless interface strength. For small values of $A/F$, the vortex ring could penetrate the interface.  As $A/F$ was increased, the interface began to act more like a solid interface, preventing penetration.

To our knowledge,  there has been no systematic study of the \textit{oblique} incidence of a vortex ring upon a density interface separating two fluids.  This is significant because the boundary conditions at a fluid density interface are different than those at a fluid/solid interface.  More generally, we lack a complete understanding of the dynamics and stability of a vortex ring propagation across a fluid density interface, as might be encountered in oceanic thermoclines or atmospheric temperature inversions.  With this in mind, we have performed a set of experiments in which we measured the trajectory of a fluid vortex ring launched at an oblique angle toward a fluid density interface.  

The present article, which describes the results of our experiments, is organized as follows.  We begin in Sec.~\ref{sec:dimensionless} by providing definitions of a number of dimensionless quantities which will prove useful in characterizing our results.  Next, in Sec.~\ref{sec:apparatus}, we describe the experimental procedure.  This section includes details concerning the vortex launcher design and operation, establishment and characterization of the fluid density interface, and the vortex tracking and image analysis techniques.  In Sec.~\ref{sec:no_interface}, we describe some control experiments in which we track the trajectory of a vortex ring in the absence of an established fluid density interface.  We do this in order to identify vortex ring instabilities which may give rise to spontaneous changes in vortex ring trajectory.  Having done this, we next report, in Sec.~\ref{sec:yes_interface}, experiments in which we track the trajectory of a vortex ring when it strikes a fluid density interface. In particular, we describe the experimental conditions under which we observed vortex ring reflection and refraction.  The experimental results are collected in Tab.~\ref{tab02}.  Finally, in Sec.~\ref{sec:discussion} we speculate about how one might construct a generalized law of refraction, akin to Snell's law for the refraction of light, in order to characterize the interaction of a vortex ring with a fluid density interface.

\section{Dimensionless quantities}
\label{sec:dimensionless}

In order to characterize our results, we will make use of a number of dimensionless quantities.  The Atwood number,  $A  =  ({\rho_2 - \rho_1})/({\rho_2 + \rho_1})$, is a dimensionless quantity which measures the density difference across an interface between two fluids.  Here, $\rho_1$ and $\rho_2$ are the densities of the fluids on the top and bottom sides of the interface, respectively.  The length scale, $l/w$, is a dimensionless quantity formed from the interface thickness, $l$, and the vortex ring diameter, $w$.  The Froude number, $F$, is a dimensionless quantity which measures the relative importance of the inertial and gravitational effects on the flow of a mass of fluid.  It has several formulations.  We will use the one provided by~\citet*[][]{landau:1987}, $F = {v}/{\sqrt{wg}}$. Here, $w$ and $v$ are the diameter and speed of the vortex ring relative to the surrounding fluid, respectively, and $g$ is the acceleration of gravity.  The diameter of the vortex ring is defined as the distance between the center of the vortex core regions on opposite sides of the ring.  The Reynolds number, $R = w v / \nu$, is defined as the ratio formed from the product of the width and velocity of the vortex ring and the kinematic viscosity, $\nu$, of the ambient fluid.  In addition, we can combine the Froude and Atwood numbers into a single dimensionless quantity, ${A}/{F} = \Delta \rho \sqrt{wg}/2  \rho_{a} v$, which plays the role of the interface strength.  Here, $\rho_{a} = \left(\rho_2 + \rho_1\right) /2$ is the average fluid density, and $\Delta \rho = \left( \rho_2 - \rho_1\right)$ is the difference between the top and bottom fluid densities.  

For a typical set of experimental runs, we keep the Atwood number and the (initial) vortex ring width constant and vary the vortex ring velocity.  For such a set, a large vortex ring velocity corresponds to a large Froude number and hence a small interface strength.  In what follows, we will occasionally use the terms ``large vortex ring velocity", ``large Froude number" and ``small interface strength" interchangeably.

\section{Experimental Procedure}
\label{sec:apparatus}

\begin{figure}[hftbp]
\begin{center}
\includegraphics[width = 4.375in]{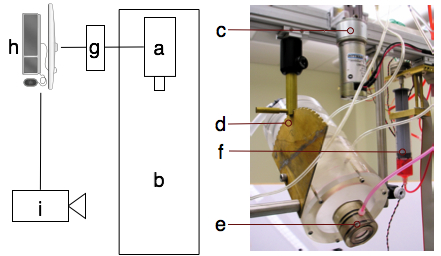}
\caption{(Color online) \textit{Left}: A schematic diagram of the experimental configuration.  \textit{Right}: A photograph of the vortex ring launcher.  See the text for identification of the labels.}
\label{fig01}
\end{center}
\end{figure}

The left hand side of Fig.~\ref{fig01} shows a schematic diagram of the experimental configuration; the right hand side is a detailed photograph of the vortex ring launcher and some of its auxiliary equipment.   The design of our vortex ring launcher was largely inspired by that of~\citet*{Lim:1989p1676}.  Briefly, the vortex ring launcher (a) was suspended inside of a 30$''$L x 12$''$W x 22$''$D glass aquarium (b) by an adjustable platform.  The platform height could be controlled with a screw jack driven by a servo motor (c).  The declination of the launcher could be adjusted manually using a notched protractor machined from marine brass (d).   

Vortex rings were ejected from the barrel of the launcher (e) by one of  two methods.  In the first method, a stepper motor sealed in a water tight chamber in the rear of the vortex launcher was used to actuate a rigid acrylic piston so as to force a precisely controlled volume of fluid through the barrel. In the second method, the acrylic piston was replaced by a latex diaphragm; the diaphragm was actuated pneumatically by quickly squeezing a pipet bulb so as to force air into the water tight chamber behind the diaphragm.  The second method had the capability of more rapid actuation and thus higher velocity vortex rings; the first method had the virtue of greater reproducibility.  

In either case, a vortex ring was formed which propagated at a particular angle and speed toward the fluid density interface.  Generally speaking, the initial vortex ring speed was limited by both the maximum speed of the actuator employed and the ratio of the cross sectional areas of the barrel and the piston (or diaphragm).  The diameter of the barrel, $d$, determined the initial diameter of the vortex ring.  In our experiments, we used barrels with two different bores: the smaller had $d=0.635$ cm, and the larger had $d= 1.905$ cm. 

Just prior to actuation of the piston, a small quantity of neutrally buoyant red ink was injected into the barrel using a syringe (f) depressed by a linear actuator.  The ink injection system (and the rigid piston, when using the stepper-motor method) were controlled using a multifunction data acquisition system (g).  Experimental control software ran on a desktop computer (h).  Concurrently running video capture software was used to obtain movies of the vortex ring trajectory using a high-definition digital video camcorder (i).  The image capture rate was 30 frames per second.

\subsection{Density Interface}

To establish a fluid density interface, we used a procedure similar to the one described by~\citet*[][]{Dahm:1989p8}.  The lower fluid layer consisted of a mixture of deionized water and salt, the upper layer of pure deionized water.  As mentioned in Sec.~\ref{sec:dimensionless}, the density difference between the fluid layers dictates the Atwood number of the interface.  In our experiments, the Atwood number was varied between 0.00147 and 0.00645. 

\begin{figure}[hftbp]
\begin{center}
\includegraphics[width = 5.375in]{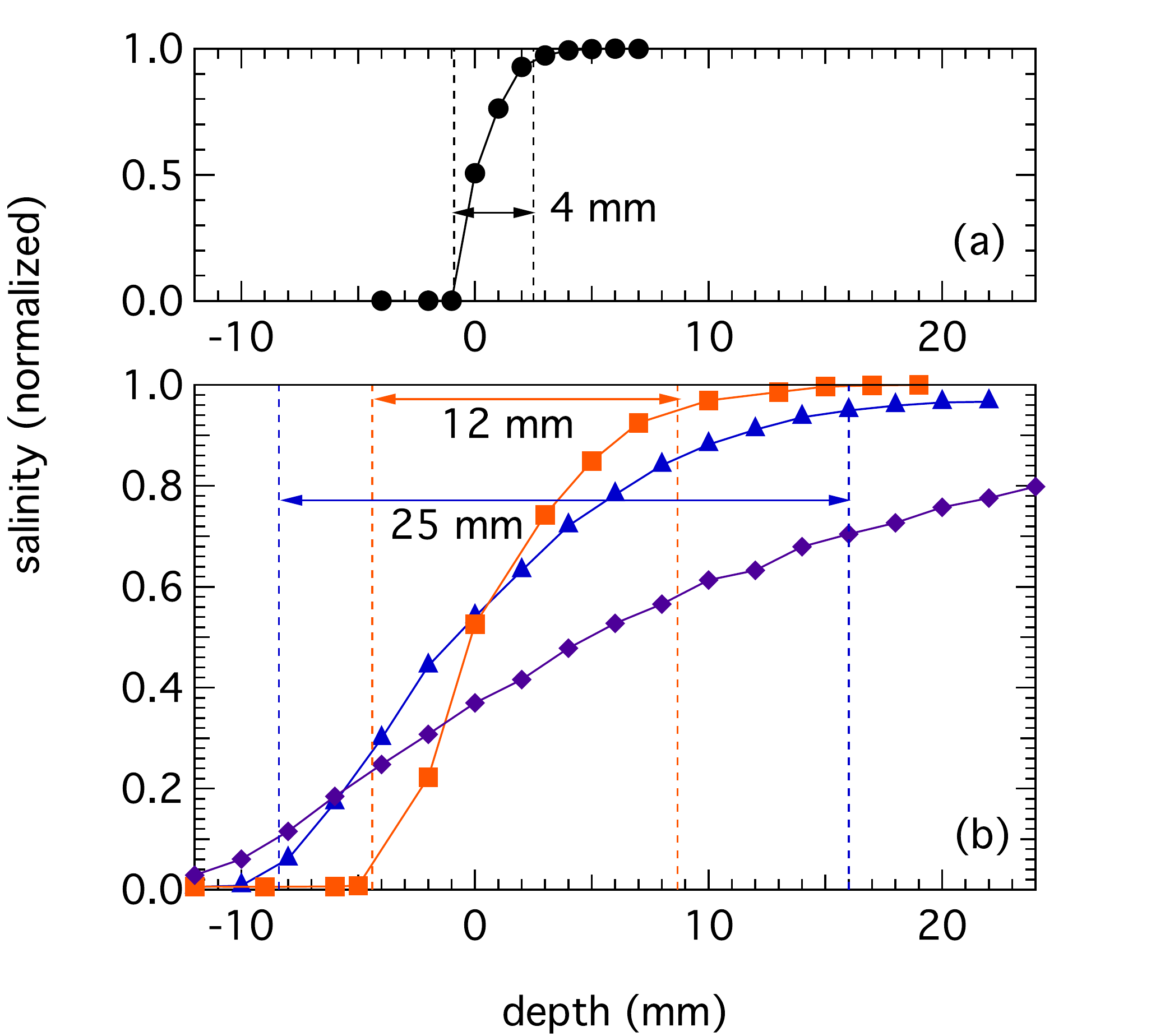}
\caption{(Color online) (a) The normalized salinity profile of an air/saltwater interface.  The measured interface width is 4 mm. The maximum salinity in (a) corresponds to a density of 1.00748 g/cc.  (b) The normalized salinity profile of a freshwater/saltwater interface immediately after it was established (orange squares), after disruption by vortex launches (blue triangles), and after allowing diffusion for six days (purple lozenges).  For the squares and triangles, the measured interface widths are 12 and 25 mm, respectively. The maximum salinity in (b) corresponds to a density of 1.00339 g/cc.}
\label{fig02}
\end{center}
\end{figure}

To measure the interface thickness, we incrementally lowered the tip of a commercially available salinity sensor across the interface.  A normalized salinity profile of the interface thus obtained is shown in Fig.~\ref{fig02}.  In the top graph (a), the salinity profile of a saltwater/air interface is shown.  The maximum salinity here corresponds to a fluid density of 1.00748 g/cc.  The interface thickness was measured to be 4 mm.  This width is taken to be the instrumental broadening due to the limited resolution of the probe.  

In the bottom graph (b), the salinity profile of a saltwater/freshwater interface is shown.   The maximum salinity corresponds to a fluid density of 1.00339 g/cc.  There are three traces.  The orange squares show the profile immediately after it was established.  The interface has a thickness of 12 mm.  The blue triangles show the profile after several vortex ring impacts.  The interface has widened to 25 mm.  The purple lozenges show the profile after the saltwater has been allowed to diffuse into the freshwater for six days, without any further vortex ring impacts.  The interface is now over 30 mm wide.  For all of these measurements, we define the interface thickness as the height of the region over which the salinity differed by more than 5\% from its maximum, and its minimum concentrations.  

To avoid excessive interface broadening during the course of our experiments, we routinely drained the tank and reestablished an interface after three to five launches. Correcting for the instrumental broadening, the average interface thickness was assumed to be $l=1.5\pm0.6$ cm.

\subsection{Image processing}

Movies of the vortex ring trajectories were analyzed with the aid of a commercially available image processing software package.  The central goal of image processing was to identify the location, size, and orientation of the vortex ring in each movie frame.  To this end, either an automated or a manual vortex tracking procedure was employed.  These two procedures are described next.

\subsubsection{Automated vortex tracking procedure}

When employing the automated vortex tracking procedure, first, a background image, which typically consisted of the first movie frame (prior to vortex launch) was subtracted from the subsequent movie frames.   This minimized the effect of spurious background features on vortex ring identification.  Next, the first two frames which contained vortex rings were inspected by eye, and the cartesian coordinates of the center of the vortex ring in each of these two frames was estimated.  From these coordinates, the initial velocity of the vortex ring was calculated.  The velocity was then used to predict a region of interest for identifying the vortex ring in the subsequent frame. Software then identified the location of contiguous image pixels whose red value (the color of the dye marking the vortex ring) exceeded a certain threshold value.  An ellipse was fit to this region, identifying the outer contour of the vortex ring.  Various vortex ring parameters such as its major axis, its orientation and its center were calculated from the properties of the ellipse.  Once the coordinates of the vortex ring were found, the velocity was re-calculated, and a region of interest was predicted for the next frame.  This procedure was iterated until the coordinates of the vortex ring in each movie frame were identified.

The virtue of this automated procedure was its speed.  In many cases, a 300-frame movie could be analyzed in a matter of seconds.  There were, however, problems with this procedure.  In particular, the code was occasionally unable to identify the vortex ring without manually adjusting the red threshold for each frame.  This was necessary because the vortex ring would sometimes fade as the red ink was shed during the vortex ring flight.  We will have more to say about vorticity shedding in Sec.~\ref{sec:no_interface}.  In any case, having to continually adjust the red threshold greatly slowed down the automated vortex tracking procedure.  But the main problem created by manually adjusting the red threshold was that it necessarily introduced systematic errors: changing the threshold changed the calculated size of the vortex ring, which affected the calculated Reynolds and Froude numbers of the flow.  These problems necessitated a more reliable tracking procedure for all but the most straightforward cases.

\subsubsection{Manual vortex tracking procedure}

Although it was much slower, a manual vortex tracking procedure was generally employed.  As in the automated vortex tracking procedure, a background image was first subtracted from each movie frame.   Next, for each frame, the cartesian coordinates of the top and the bottom core regions of the vortex ring were identified by eye.  This was easily done because the red dye tended to be most concentrated in the core regions.  From these coordinates, the width, orientation, and center of the vortex ring were calculated.  

Although this procedure was certainly more tedious, it enjoyed a number of advantages over the automated tracking procedure.  First, it was more reliable, in the sense that there was far less scatter in the calculated width of the ring during its flight.  Second, it was a straightforward matter to assign an experimental uncertainty to the coordinates of the vortex ring by estimating the precision with which the centers of the vortex core region could be identified.  Finally, the identification of the width of the vortex ring with the distance between the top and bottom core regions, rather than the extent of the visible dye marking the entire region of the vortex ring, has the advantage of being consistent with the conventional way of measuring vortex ring width, $w$, employed by previous workers. 

\begin{figure}[hftbp]
\begin{center}
\includegraphics[width=5.375 in]{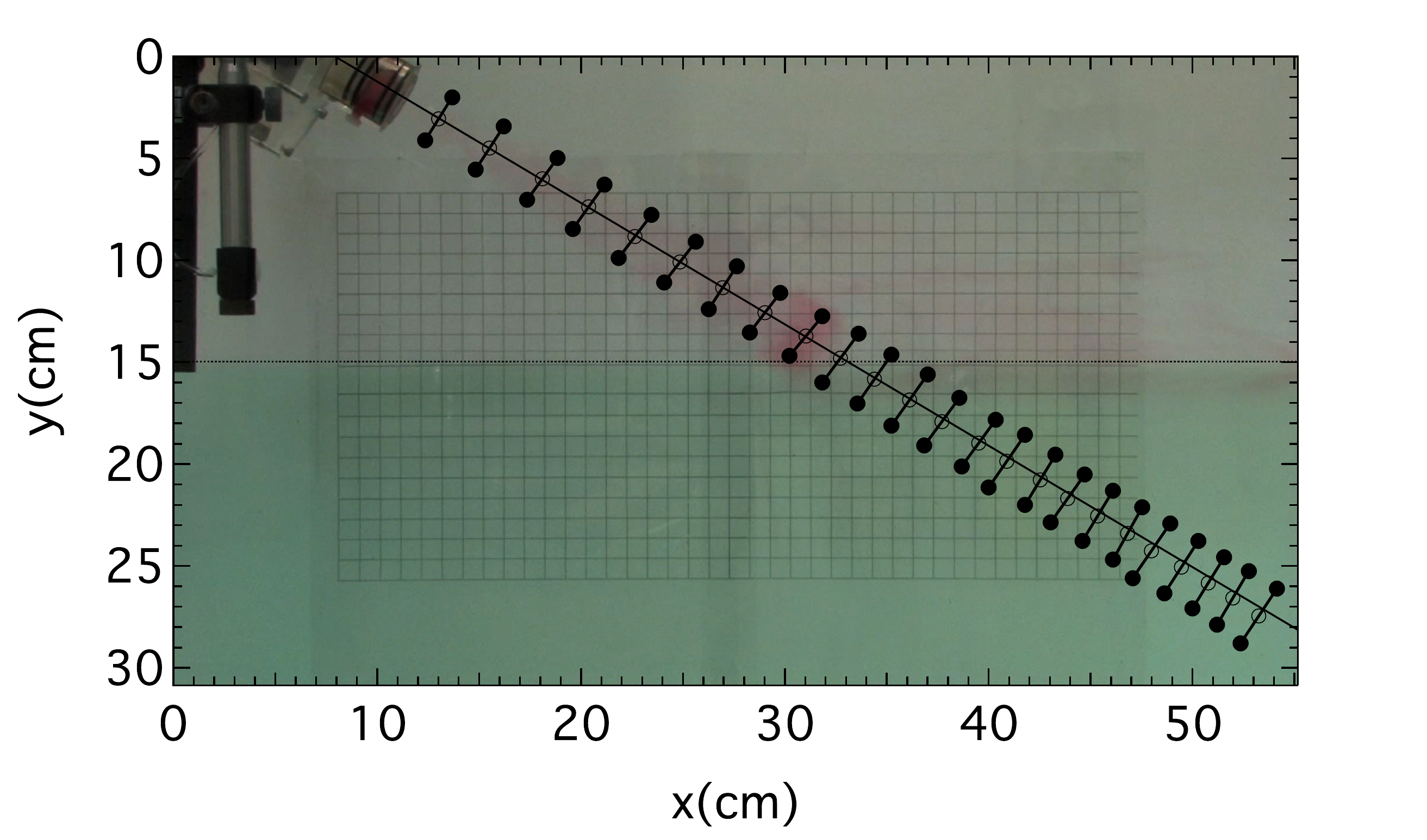} 		
\caption{(Color online) A montage depicting the position, and orientation, of a vortex ring launched from the upper left corner of the image at an angle of 59.2 degrees with respect to the vertical.  The horizontal dotted line indicates the position of the center of the fluid density interface. The oblique black line indicates the initial trajectory of the vortex ring. The Atwood number for this interface is $A=0.00428$. }
\label{fig03}
\end{center}
\end{figure}

The results of using the manual vortex tracking procedure are illustrated in Fig.~\ref{fig03}.  This figure is a montage depicting the location and orientation of a vortex ring launched from the upper left corner of the image at an angle of 59.2 degrees with respect to the vertical.  In the background image selected for this figure, a faintly visible vortex ring (marked with red dye, which can be more clearly seen in the color online version) is about to strike the interface, near the coordinates $(x,y) = (30, 14)$ cm.  The 1.0 square centimeter grid, visible on the back wall of the water tank, was used for length scale calibration.  Also visible in the background image are the barrel of the launcher and the vertically oriented salinity probe.

Superimposed on the background image: the closed circles indicate the top and bottom cores of the vortex ring; the open circles indicate the center of the vortex ring; and the connecting lines serve as a guide to the eye for discerning the vortex ring orientation.  The horizontal dotted line indicates the position of the center of the fluid density interface. In this case, the saltwater has been marked with a green dye to make the interface more distinct.  The solid line is a linear fit to the data points above the fluid density interface.  The angle between this line and the vertical is used to compute the incidence angle of the vortex ring trajectory. Notice that the trajectory of the vortex ring deviates slightly downward from this fit line after penetrating the interface.  We will have more to say about this when we discuss vortex ring refraction in Sec.~\ref{sec:yes_interface}.

\begin{figure}[hftbp]
\begin{center}
\includegraphics[width=3.6in]{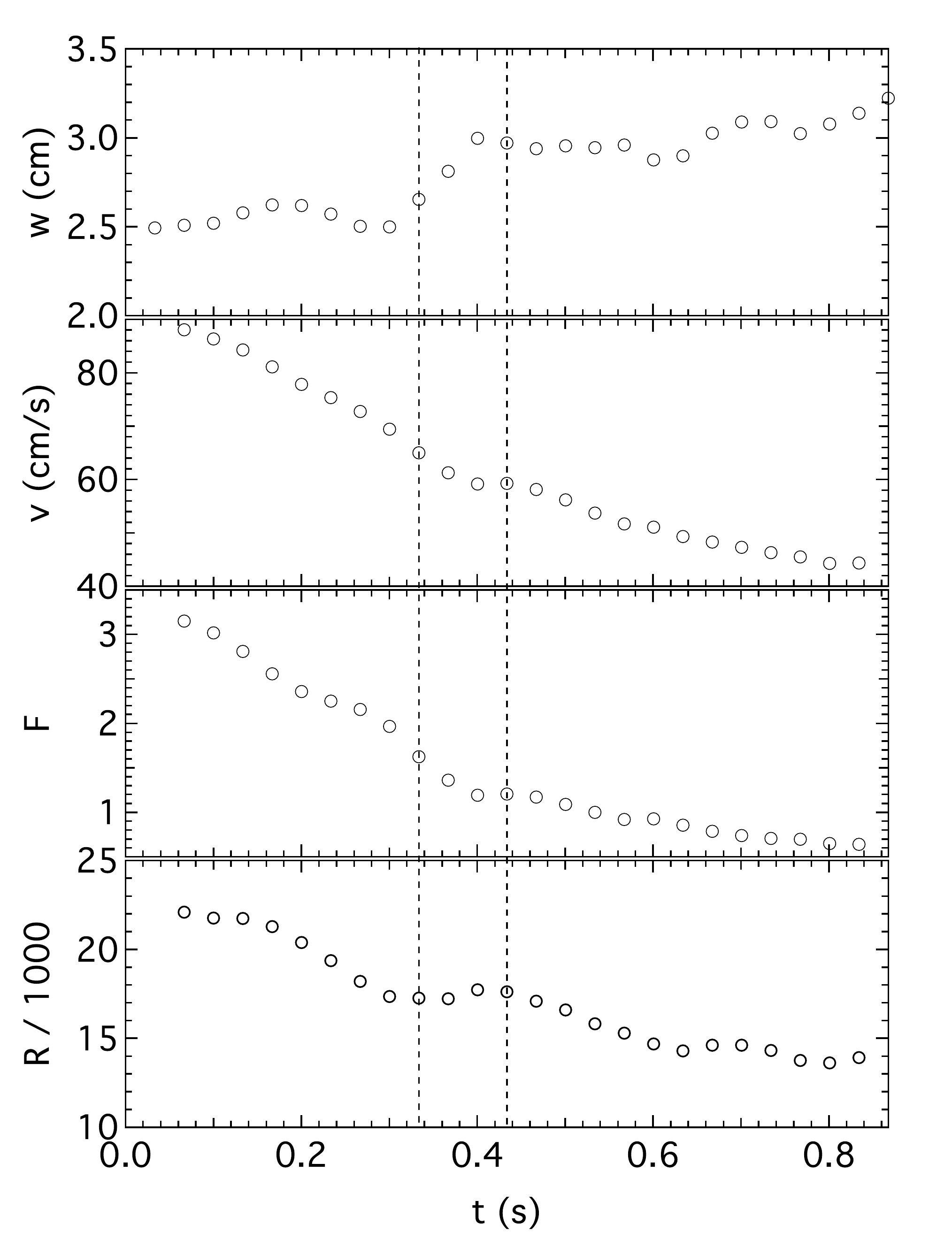} 	
\caption{The time dependence of the width, $w$, velocity, $v$, Froude number, $F$, and Reynolds number, $R$, for the launch depicted in Fig.~\ref{fig03}.  The vertical dashed lines delimit the time interval during which the vortex ring is in contact with the fluid density interface.}	
\label{fig04}
\end{center}
\end{figure}

From the data shown in Fig.~\ref{fig03}, one can compute the width, velocity, Froude number and Reynolds number of the vortex ring during its flight.  These values are shown in Fig.~\ref{fig04}.  A boxcar average with a width of 3 data points was empolyed for each trace.  In this figure, the vertical dashed lines delimit the time interval during which the vortex ring was in contact with the center of the fluid density interface.  Notable in Fig.~\ref{fig04} is an abrupt change in the vortex ring width, $w$, which occurs while the vortex ring is traversing the interface.  This change in width is accompanied by a marked change in the slope of the velocity, $v$, and the Froude number, $F$.  Since the Reynolds number, $R$, is derived from the product of the velocity and width, its change due to the impact is somewhat less pronounced.  Notice, too, that the Reynolds number has been divided by 1000 in the bottom graph of Fig.~\ref{fig04}.

\section{Vortex trajectories in the absence of a density interface}
\label{sec:no_interface}

\begin{figure}[hftbp]
\begin{center}
\includegraphics[width=5.375in]{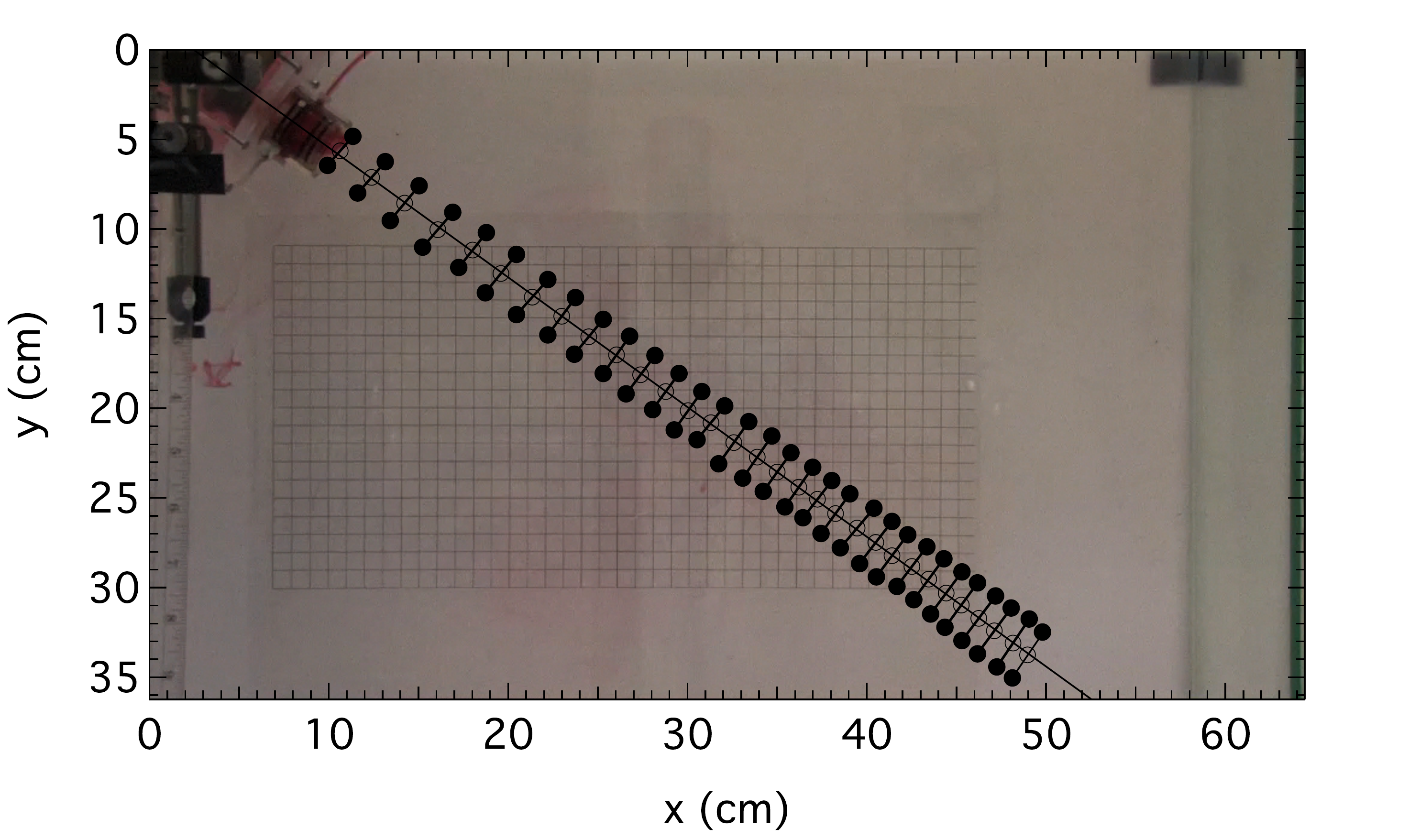}		
\caption{(Color online) A montage depicting the position, and orientation, of a vortex ring launched from the upper left corner of the image at an angle of 54.1 degrees with respect to the vertical. There is no fluid density interface; the trajectory is linear.}	
\label{fig05}
\end{center}
\end{figure}

\begin{figure}[hftbp]
\begin{center}
\includegraphics[width=3.6in]{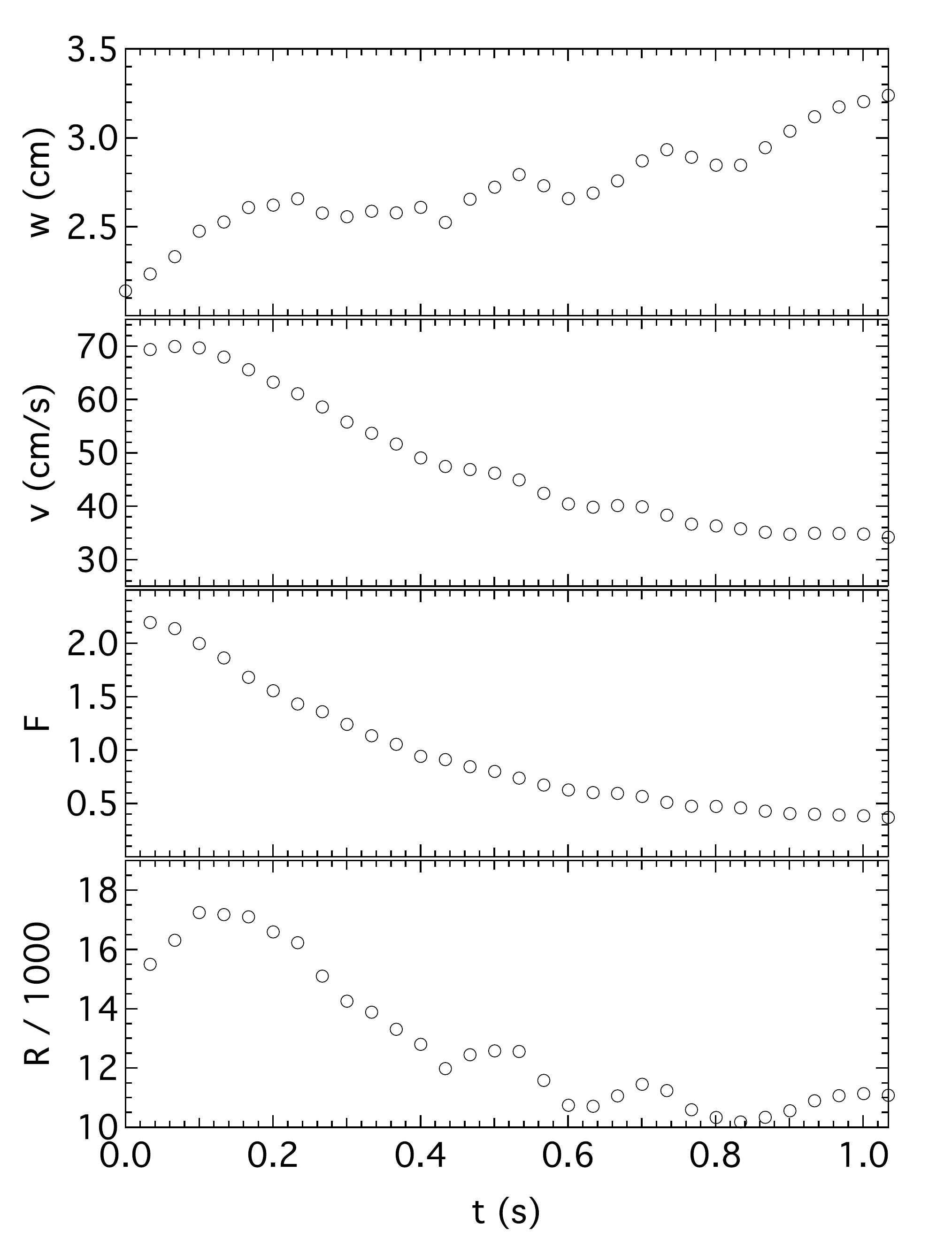}			
\caption{(Color online) The time dependence of the width, $w$, velocity, $v$, Froude number, $F$, and Reynolds number, $R$, for the launch depicted in Fig.~\ref{fig05}.}
\label{fig06}
\end{center}
\end{figure}

Before describing in more detail our experiments in which a vortex ring strikes a fluid density interface at an oblique angle,  we describe briefly the behavior of a vortex ring propagating through a fluid in the absence of a fluid density interface.  Such a case is illustrated in Fig.~\ref{fig05}.  This is a montage depicting the location and orientation of a vortex ring launched from the upper left corner of the image at an angle of 54.1 degrees with respect to the vertical.  It is clear that in Fig.~\ref{fig05}, the trajectory is linear.  Moreover, as shown in Fig.~\ref{fig06}, the width and velocity, suffer no abrupt changes.  The Froude number, computed from the ratio of the velocity and width, is a particularly smooth function.  The observed variations in the Reynolds number are due to variations in measurement of the vortex ring width.

\subsection{Spontaneous instabilities}
\label{subsec:instabiliites}

Such smooth behavior is not always the case.  In certain launches, the vortex ring width has been observed to change spontaneously and abruptly.  This is typically accompanied by an abrupt change in the slope of $v(t)$, and the shedding of vorticity.  The shedding of vorticity is manifested either as a deposition of a significant quantity of dye into the wake of the vortex ring, or as the spawning of a new, smaller, vortex ring whose trajectory is oblique to that of the parent vortex ring. In the latter case, the trajectory of the parent vortex ring changes, as might be expected from the conservation of momentum.  

\begin{figure}[hftbp]
\begin{center}
\includegraphics[width=5.375in]{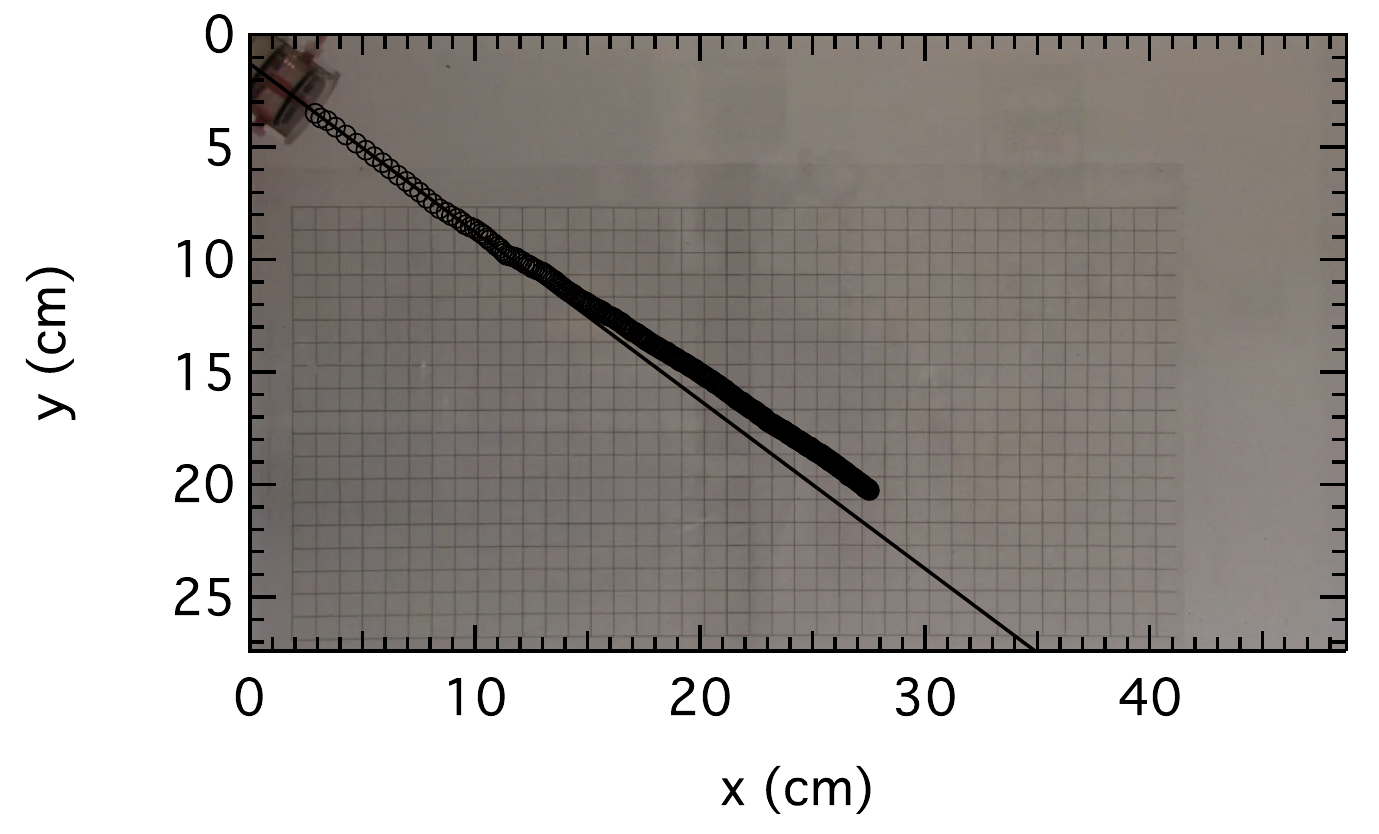}			
\caption{(Color online) A montage depicting the position of a vortex ring launched from the upper left corner of the image at an angle of 53.2 degrees with respect to the vertical. There is no fluid density interface. The vortex ring clearly suffers a change from its initial trajectory, indicated by the oblique black line, due to spontaneous shedding of vorticity (not shown).} 
\label{fig07}
\end{center}
\end{figure}

Consider, for example, the launch depicted in Fig.~\ref{fig07}.  Due to the density of data points in this launch, we have omitted the vortex orientation lines for the sake of clarity.  Midway through its flight, the vortex ring was observed to suddenly shed a small quantity of vorticity downward (not shown).  Simultaneously, its trajectory deflected upwards.  

\begin{figure}[hftbp]
\begin{center}
\includegraphics[width=3.6in]{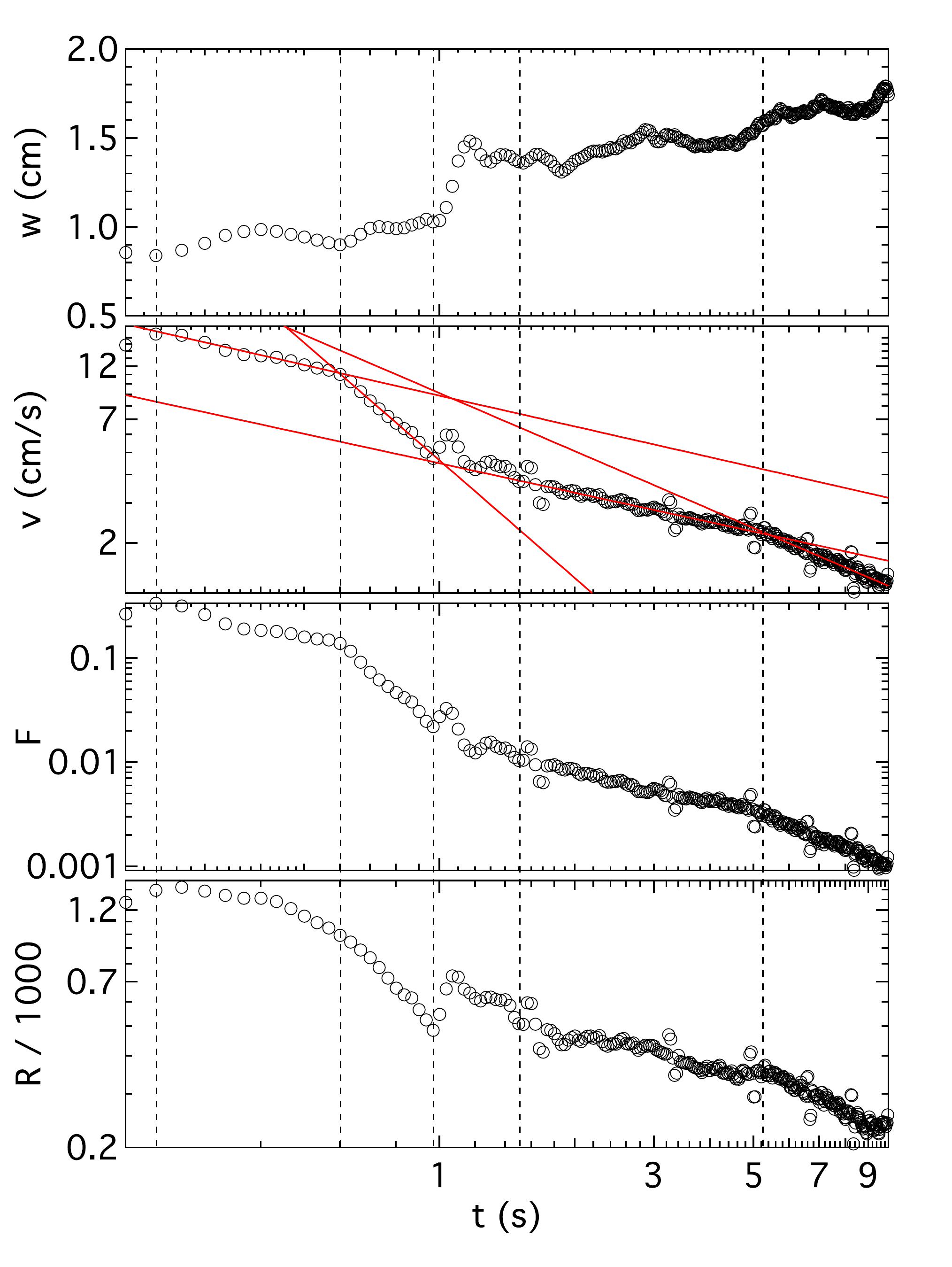}			
\caption{(Color online) The time dependence of the width, $w$, velocity, $v$, Froude number, $F$, and Reynolds number, $R$, for the launch depicted in Fig.~\ref{fig07}.  The time of instabilities are marked by vertical dashed lines.  Power law fits for each stable region are shown as oblique solid red lines in the velocity plot.}
\label{fig08}
\end{center}
\end{figure}

Upon inspecting the time dependence of the vortex width, shown in the top graph of Fig.~\ref{fig08}, it was found that at the moment the vortex ring trajectory changed, the vortex ring experienced an abrupt 30\% increase in its width.  This occurred at $t=0.97$ s.  At the same time, the Reynolds number fell below 600, as shown in the bottom graph.  In Fig.~\ref{fig08}, a logarithmic scale has been used on the abscissa, as well as on the ordinate for the velocity, Froude number and Reynolds numbers.

This is reminiscent of the observations of~\citet*[][]{Maxworthy:1972p15}, who found that a vortex ring tends to become unstable with respect to the formation of a new, stably propagating vortex ring when the Reynolds number is between 600 and 1000. Maxworthy attributed this instability to the growth of azimuthal waves on the vortex ring.  

On the other hand, Maxworthy also observed that when $R \lesssim 600$, a vortex ring is ``stable during the whole of its motion."  Our data suggests that this is not necessarily the case.  Upon closer examination of the $v(t)$ graph in Fig.~\ref{fig08}, there appear to be a number of instability events.  These instabilities are identified by abrupt changes in the slope of $v(t)$. The times at which the slope of $v(t)$ changes are indicated by five vertical dashed lines at times $t_i =$0.27, 0.60, 0.97, 1.47 and 5.24 seconds. 

During all of the time intervals (except for the transient region during the shedding between 0.97 and 1.47 seconds) the velocity of the vortex ring appeared to vary linearly on a log-log plot.  We thus performed a power law fit, $v(t) = A t^{-\alpha}$, to the data during each of the stable time intervals.  These fits are shown as oblique solid red lines on the graph of $v(t)$ in Fig.~\ref{fig08}.   A list of the observed instabilities, the Reynolds numbers at which they occur, and the power law fitting parameters in the stable time intervals, are shown in Tab.~\ref{tab01}.  The uncertainties in the fitting parameters reported in Tab.~\ref{tab01} are computed assuming a normal distribution of measurement errors. The uncertainties in the Reynolds numbers reported in Tab.~\ref{tab01}, obtained from Fig.~\ref{fig08}, are approximately 10\%.  

It should be noted that the last observed instability for this launch occurred at $t = 5.24$ s, when $R = 350$.  This is in the range of $R$ where Maxworthy observed stable vortex ring propagation. It should also be mentioned that an alternative fit, of the form $v(t) = A (t-t_0)^{-3/4}$, based on the assumption of self-similar vortex ring propagation~\cite{Glezer:1990p243}, yielded consistently larger values of $\chi^2$, and hence a poorer fit to the data, than the power law fit that we finally adopted. 

\begin{table}[hftbp]
\caption{Summary of spontaneous vortex ring instabilities occurring during the launch depicted in Fig.~\ref{fig07}.  The $t_i$ demarcate the regions of stable propagation. $A$ and $\alpha$ are fitting parameters for a power law fit, $v(t) = A t^{-\alpha}$, to the data between time $t_i$ and $t_{i+1}$.  $R$ refers to the Reynolds number at $t_i$. Uncertainties in the fitting parameters are also shown.}
\begin{center}
\begin{tabular}{|D..{2.3}|D..{2.3}D..{1.3}|D..{2.3}D..{1.3}|D..{5.1}|}
\multicolumn{1}{|c|}{$t_i$} & 
\multicolumn{2}{c|}{$A$} & 
\multicolumn{2}{c|}{$\alpha$} & 
\multicolumn{1}{c|}{$R$}  \\
\hline
\hline
0.27 & 8.89 & \pm0.23 & 0.45 & \pm0.03 & 1430 \\
0.60 & 4.61 & 0.06 & 1.72 & 0.03 & 990 \\
0.97 &   &  &  & & 566 \\
1.47 & 4.48 & 0.08 & 0.43 & 0.02 & 530 \\
5.24 & 9.13 & 0.40 & 0.85 & 0.02 & 350 \\
\hline \hline
\end{tabular}
\end{center}
\label{tab01}
\end{table}

A great deal has been accomplished by other workers in trying to identify the particular mechanism(s) which gives rise to vortex ring instability in various situations~\cite[][]{Maxworthy:1977p465}.  By contrast, we have not attempted to analyze the internal structure of the vortex ring.  Our analysis is based strictly on parameters such as the width, velocity, Froude number, Reynolds number and trajectory of the vortex ring.  The important point, for purposes of what follows, is that in the absence of manifest oblique vorticity shedding, the vortex ring appears to follow a linear trajectory.  We will describe in the following section how the presence of a fluid density interface, on the other hand, gives rise to obvious and consistent deviations from a linear trajectory.

\section{Vortex trajectories in the presence of a density interface}
\label{sec:yes_interface}

We now consider the results of a number of experiments performed in which a vortex ring was launched at an oblique angle toward a fluid density interface.   In order to illustrate the most significant results, we have divided our presentation into two sections, the first for large angles of incidence ($\theta_i > 70$ degrees), the second for small angles of incidence  ($\theta_i < 70$ degrees).  In each of the trajectory plots that follow, we only show the trajectory of the center of the vortex ring; we omit the vortex orientation lines as well as the background images for the sake of clarity.  A summary of the experimental parameters for each of the launches in this section is provided in Tab.~\ref{tab02}.

\subsection{Large incidence angle: reflection at large interface strength}

\begin{figure}[hftbp]
\begin{center}
\includegraphics[width=6.375in]{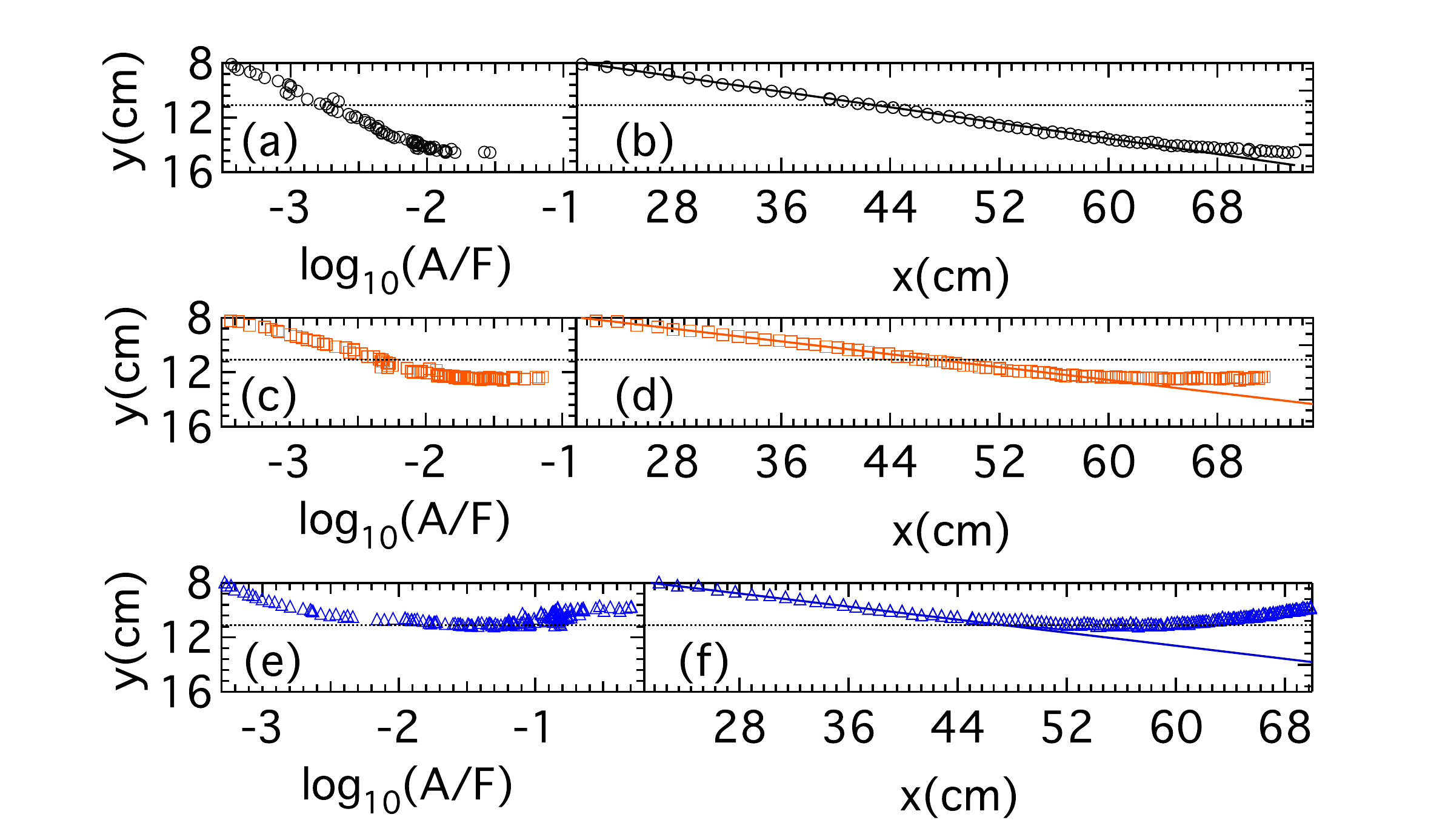}				
\caption{(Color online)  Data for three separate vortex ring launches at an incidence angle of 82 degrees with respect to a normal to the fluid density interface, shown as a horizontal dotted line. In the right column, for each launch, is shown the trajectory, as well as a linear fit to the first ten to twenty data points above the interface; in the left column is shown the depth dependence of $\log_{10}({A/F})$.}	
\label{fig09}
\end{center}
\end{figure}

Fig.~\ref{fig09} depicts the results of three launches, each at an incidence angle of 82 degrees with respect to the vertical.  In the top row (black circles) are shown (b) the trajectory and (a) the depth dependence of the logarithm of the interface strength for a vortex launched at a relatively large velocity towards the fluid density interface.  A large velocity corresponds to a small value of $\log_{10}(A/F)$.  The interface location is indicated by the horizontal dotted line near $y = 11$ cm.  For this launch, the vortex ring is able to penetrate the interface, but it clearly deviates from its linear trajectory after entering the higher density fluid.  The upward curvature can be understood as a consequence of buoyancy, since after penetrating the interface, the vortex is comprised of a volume of low density fluid immersed in an ambient high density fluid. 

The second row (orange squares) of Fig.~\ref{fig09} shows another launch at an incidence angle of 82 degrees.  In this case, however, the vortex ring speed is slightly lower.  Consequently, the vortex ring does not penetrate as deeply into the bottom fluid before the effect of buoyancy drives it back upwards, towards the interface.  In the third row (blue triangles), the speed of the ring is even slower.  In this case, the vortex ring is unable to penetrate the interface.  Instead, it simply reflects from the interface.  The trend displayed in Fig.~\ref{fig09} is consistently observed:  the slower moving vortices, characterized by the larger values of $\log_{10}(A/F)$ at impact, tend to reflect from a fluid density interface, provided the incidence angle is sufficiently large.

\begin{figure}[hftbp]
\begin{center}
\includegraphics[width=6in]{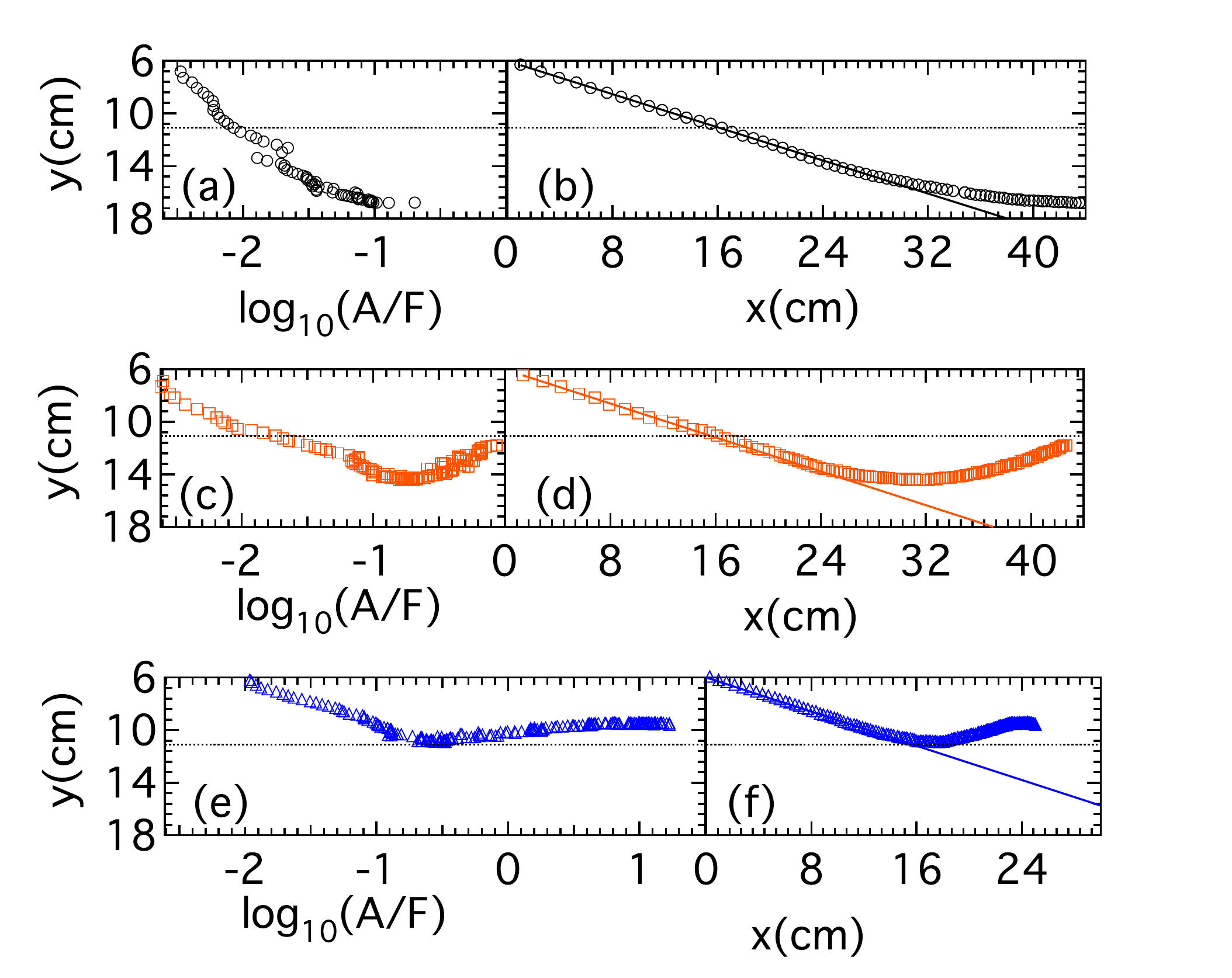}				
\caption{(Color online)  Data for three separate vortex rings launches at an incidence angle of 72 degrees with respect to a normal to the fluid density interface, shown as a horizontal dotted line. In the right column, for each launch, is shown the trajectory, as well as a linear fit to the first ten to twenty data points above the interface; in the left column is shown the depth dependence of $\log_{10}({A/F})$.}	
\label{fig10}
\end{center}
\end{figure}

Fig.~\ref{fig10} depicts the results of three more launches.  In this case, the incidence angle is only 72 degrees with respect to the vertical.  Nonetheless, the results are similar to those depicted in Fig.~\ref{fig09}. The rapidly moving vortex shown in the top row (black circles) penetrates the interface before experiencing an upward curvature due to buoyancy.  As the speed is decreased (orange squares), the vortex ring does not penetrate as deeply until, at some threshold value, it merely reflects off of the interface, as shown in the bottom row (blue triangles).  

Interestingly, for the launch shown in the bottom row of Fig.~\ref{fig10}, the reflected vortex exhibits a cycle of damped oscillations prior to disintegrating.  This is likely due to entrainment of the surrounding fluid, which leads to alternating upward and downward forces acting on the ring during its flight.  A similar effect was observed by~\citet*[][]{Scase:2006p302} when studying the propagation of vortex rings through a stably stratified fluid.

\subsection{Small incidence angles: refraction at small interface strength}

\begin{figure}[hftbp]
\begin{center}
\includegraphics[width= 6 in]{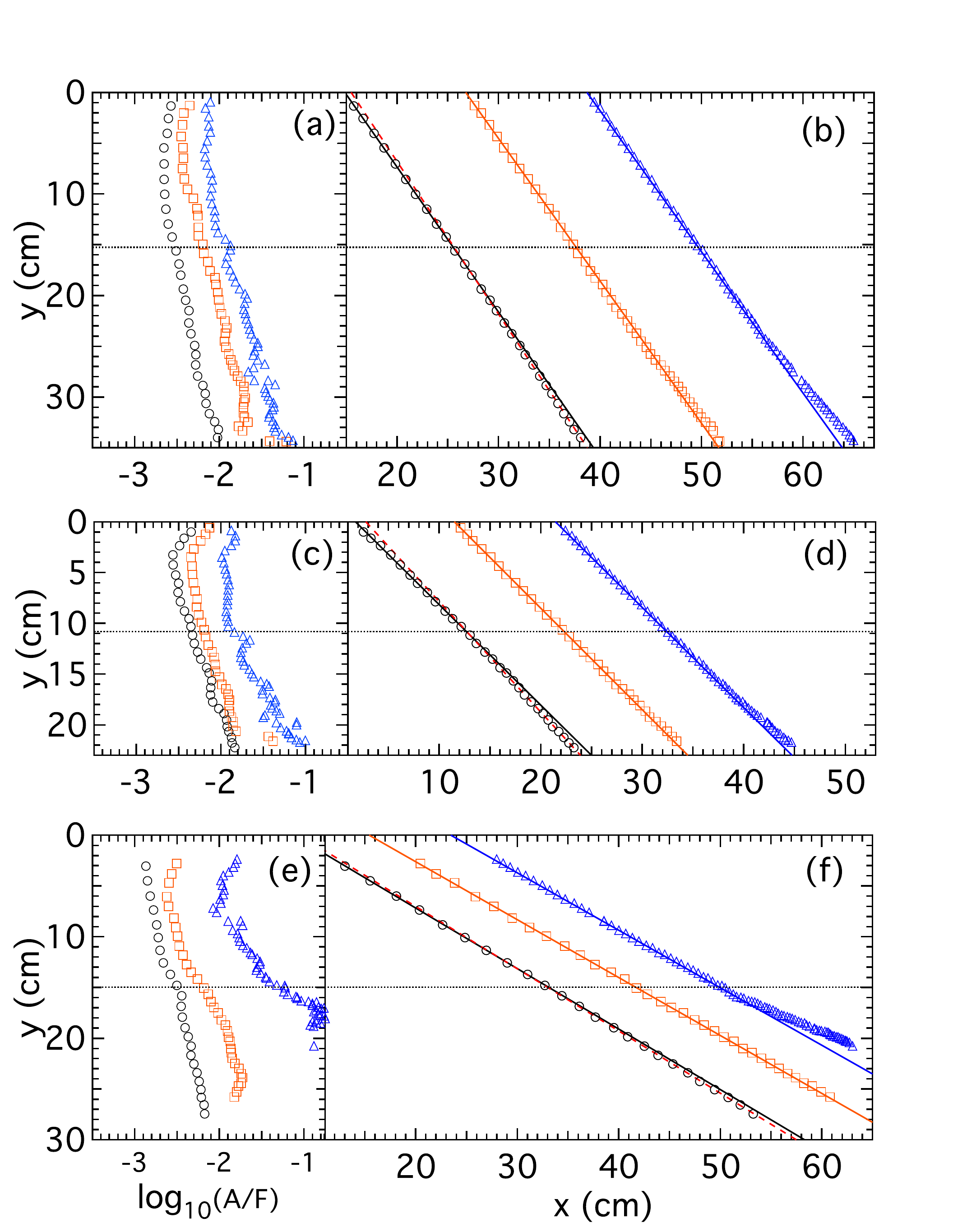}			
\caption{(Color online)  Data for vortex rings launched at incidence angles 35 (top row), 45 (middle row) and 60 degrees (bottom row). In each row, three separate launches are shown, represented from right to left by blue triangles, orange squares and black circles.  The trajectories are shown in the right column, along with linear fits to the data above the interface (solid lines) and below the interface (dashed red lines).  The depth dependence of $\log_{10}({A/F})$ is shown in the left column.}	
\label{fig11}
\end{center}
\end{figure}

Now we present the data for launches with smaller angles of incidence. A total of nine launches are depicted in Fig.~\ref{fig11}, three in each row.  Consider, first, the right hand side of the top row (b).  Here are three launches, each having an incidence angle of 35 degrees, but each having a somewhat different speed.  The rightmost trajectory, represented by blue triangles, depicts the slowest moving vortex ring.  The middle trajectory, represented by orange squares, depicts a slightly faster vortex ring.  And the leftmost trajectory, represented by black circles, depicts the most rapidly moving vortex ring of the three.  Due to the large vortex ring speed, the interface strength is correspondingly weakest for the leftmost trajectory.  This is illustrated in the left hand side of the top row (a), where the depth dependence of the interface strength for each of these three launches is shown.

What is perhaps most significant is the fact that, while all three vortex rings follow approximately linear trajectories as they cross the fluid density interface, the most slowly moving vortex ring curves upward after striking the interface, while the most rapidly moving vortex ring refracts downward, away from the interface.  The intermediate speed vortex ring retains a linear trajectory for the duration of its flight. For each launch, the precise value of the angle of incidence,  $\theta_i$ was determined by fitting a line to the data which lies above the interface.  We also determined the angle of refraction, $\theta_r$, by fitting a line to the data which lies below the interface.  Generally speaking, this latter fit is quite good for the fast moving vortices, but less so for the slow moving vortices, due to their upward curvature after penetrating the interface.  The fits above the interface (solid lines) and below the interface (dashed red line) are shown in Fig.~\ref{fig11}(b) for the most rapidly moving vortex ring.  Careful inspection reveals that the angle of incidence is larger than the angle of refraction.

A similar tendency is revealed for the vortex rings launched at 45 degrees with respect to the vertical, shown in the middle row of Fig.~\ref{fig11}: the trajectory of the slowly moving vortex ring, represented again by blue triangles, experiences an upward curvature after penetrating the interface, whereas the trajectory of the quickly moving vortex ring, represented by black circles, experiences a downward refraction.  This phenomenon is repeated for the bottom row of Fig.~\ref{fig11}, which shows data for launches with an incidence angle of 60 degrees.

These results suggest that there are competing effects which influence the vortex ring trajectory.  For large values of the interface strength, buoyancy predominates, and the vortex ring curves upward after penetrating the interface.  For small values of the interface strength, the buoyancy effect, although presumably still present, is overcome by a force which tends to deflect the vortex ring trajectory downward, away from the interface.  Generally speaking, vortex ring refraction tends to occur when the interface strength is small.

A more detailed look at the experimental parameters characterizing the trajectory data shown in Figs.~\ref{fig09},~\ref{fig10}~and~\ref{fig11} is provided in Tab.\ref{tab02}.  The rows and columns of Tab.~\ref{tab02} are labelled so as to facilitate reference to the data contained therein.  Column 1 indicates the particular figure in which the trajectory data appears.  Here, $t$, $m$ and $b$ indicate the top, middle and bottom parts of Fig.~\ref{fig11}, respectively.  The symbols in column 1 indicate the trace for the particular trajectory: black circles, orange squares, or blue triangles. Column 2 indicates the inner diameter of the barrel, in centimeters.  Column 3 shows the dimensionless interface thickness.  Its uncertainty is dominated by the $\sim30\%$ uncertainty in the interface width, $l$.  Column 4 indicates the Atwood number.  Column 5 shows the Froude number at the moment that the vortex strikes the interface.  Column 6 provides a measure of the interface strength at the moment of impact.  Its uncertainty is estimated from the plots of $y$ versus $\log_{10}{(A/F)}$, such as those in the left hand column of Fig.~\ref{fig09},~\ref{fig10}~and~\ref{fig11}.  Column 7  indicates the Reynolds number at the moment of impact. Its uncertainty is estimated from plots of $R$ versus $t$, such as those on the bottom of Fig.~\ref{fig04},~\ref{fig06}~and~\ref{fig08}.  Columns 8 and 9 show the angles of incidence and refraction. Their uncertainties are calculated by performing linear fits to the trajectory data.  Each each fit was weighted by the estimated uncertainties in vortex ring position data.  The refraction angle is omitted from the table in rows 3 and 6 because the buoyancy induced curvature rendered it impractical to perform a reasonable linear fit to the data below the interface.  The ratio of the sines of these angles is shown in column 10.  This ratio, inspired by the law of refraction for light, provides a characterization of the vortex ring behavior when striking the interface.  For the trajectory data shown in rows 7, 10 and 13, this ratio is greater than unity.  These are the trajectories which illustrate vortex ring refraction.  Notably, for each of these data sets, $\log_{10}{(A/F)}< -2.38 \pm 0.05$.  That is, for weak interfaces, refraction is observed.  We will address the significance of column 11 in the next section.

\begin{turnpage}
\begin{table*}[hftbp]
\caption{Summary of the dimensionless quantities for the trajectory data shown in Figs.~\ref{fig09},~\ref{fig10}~and~\ref{fig11}.  There are fifteen rows; each row corresponds to one data set.  There are eleven columns; please refer to the text for the meaning of each.}
\begin{center}
\begin{tabular}{|c||ccc|D..{1.3}|D..{1.1}D..{2.1}|D..{1.5}|D..{1.3}|D..{2.3}D..{1.2}|D..{5.1}D..{3.0}|D..{2.4}D..{1.3}|D..{2.4}D..{1.3}|D..{1.4}D..{1.3}|D..{1.4}D..{1.3}|}

\multicolumn{1}{c||}{} & 
\multicolumn{3}{c|}{1} & 
\multicolumn{1}{c|}{2} & 
\multicolumn{2}{c|}{3} & 
\multicolumn{1}{c|}{4} & 
\multicolumn{1}{c|}{5} & 
\multicolumn{2}{c|}{6} & 
\multicolumn{2}{c|}{7} & 
\multicolumn{2}{c|}{8} & 
\multicolumn{2}{c|}{9} &
\multicolumn{2}{c|}{10} &
\multicolumn{2}{c|}{11} \\ \cline{2-21}

\multicolumn{1}{c||}{} & 
\multicolumn{3}{c|}{Fig.} & 
\multicolumn{1}{c|}{$d$(cm)} & 
\multicolumn{2}{c|}{$l/w$} & 
\multicolumn{1}{c|}{$A$} & 
\multicolumn{1}{c|}{$F$} & 
\multicolumn{2}{c|}{$\log_{10}(A/F)$} & 
\multicolumn{2}{c|}{$R$} & 
\multicolumn{2}{c|}{$\theta_i$(deg.)} & 
\multicolumn{2}{c|}{$\theta_r$(deg.)} &
\multicolumn{2}{c|}{ $\sin{\theta_i}/\sin{\theta_r}$ } &
\multicolumn{2}{c|}{ $v_i / v_r$ } \\
\hline \hline
1 & 9 & & $\varbigcirc$ & 0.635 & 1.5 & \pm0.6 & 0.00147 & 1.510  & -3.01 & \pm0.10 & 3640 & \pm200 & 81.840 & \pm0.016  & 84.115 & \pm0.006  & 0.995 & \pm0.002 & 0.934 & \pm0.046\\
2 &   &  & {\color{orange} \large $\square$} & 0.635 & 1.5 & 0.6 & 0.00147 & 0.533 & -2.56 & 0.05 & 2200 & 50 & 83.273 & 0.010  & 88.353 & 0.008 & 0.994 & 0.001 & 1.005 & 0.071 \\
3 &  &  & {\color{blue} \large $\triangle$} & 0.635 & 0.9 & 0.4 & 0.00147 & 0.024  & -1.93 & 0.20  & 1600 & 100 & 83.166 & 0.004  &  &  &  & & &\\
\hline 
4 &10 & & $\varbigcirc$ & 0.635 & 2.1 & 0.8 & 0.00645 & 1.117  & -2.24 & 0.10 & 2050 & 150 & 72.510 & 0.012 & 79.137 & 0.003  & 0.971 & 0.004 & 1.043 & 0.050\\
5 &   &  & {\color{orange}\large $\square$}  & 0.635 & 1.5 & 0.6 & 0.00645 & 0.489  & -1.88 & 0.10 & 2230 & 150 & 72.139 & 0.012  & 91.596 & 0.003 & 0.952 & 0.004 & 1.026 & 0.073 \\
6 &  &  & {\color{blue} \large $\triangle$}  & 0.635 & 1.7 & 0.7 & 0.00645 & 0.032 & -0.70 & 0.20 & 450 & 100 & 72.011 & 0.006  & &  & & & & \\
\hline
7 &11 & $t$ & $\varbigcirc$ & 1.905 & 0.6 & 0.2 & 0.00256 & 0.925   & -2.56 & 0.05 & 12750 & 200 & 34.965 & 0.010  & 33.616 & 0.006 & 1.035 & 0.017 & 1.036 & 0.027 \\
8 &   & & {\color{orange}\large $\square$}  & 1.905 & 0.7 & 0.3 & 0.00256 & 0.453  & -2.25 & 0.05 & 7450 & 250 & 35.474 & 0.009 & 37.255 & 0.005 & 0.959 & 0.013 & 0.987& 0.040 \\
9 &   & & {\color{blue} \large $\triangle$}  & 1.905 & 0.6 & 0.3 & 0.00256 & 0.265  & -2.02 & 0.10& 6170 & 250 & 35.710 & 0.008  & 39.195 & 0.004  & 0.924 & 0.011 & 0.992 & 0.058\\
\hline
10 &   & $m$ & $\varbigcirc$ & 1.905 & 0.7 & 0.3 & 0.00238 & 0.567 & -2.38 & 0.05 & 7510 & 200 & 45.331 & 0.010  & 42.604 & 0.007  & 1.051 & 0.013 & 1.045& 0.039 \\
11 &   & & {\color{orange}\large $\square$}  & 1.905 & 0.6 & 0.2 & 0.00238 & 0.403 & -2.23 & 0.05 & 7780 & 200 & 44.994 & 0.009 & 46.105 & 0.008 & 0.981 & 0.011 & 1.036 & 0.041\\
12 &   & & {\color{blue} \large $\triangle$}  & 1.905 &  0.7 & 0.3 & 0.00238 & 0.195 & -1.92 & 0.15 & 4980 & 250 & 45.372 & 0.008 & 48.087 & 0.005  & 0.956 & 0.009 & 1.018 & 0.059 \\
\hline
13 &   & $b$ & $\varbigcirc$ & 1.905 & 0.6 & 0.2 & 0.00428 & 1.623  & -2.58 & 0.05 & 17260 & 250 & 59.215 & 0.010  & 58.499 & 0.009  & 1.008 & 0.008 & 1.035 & 0.022 \\
14 &   & & {\color{orange}\large $\square$}  & 1.905 & 0.6 & 0.2 & 0.00428 & 0.922 & -2.33 & 0.05 & 12390 & 500 & 60.296 & 0.009 & 60.355 & 0.008  & 0.999 & 0.007 & 1.057 & 0.031 \\
15 &   & & {\color{blue} \large $\triangle$}  & 1.905 & 0.7 & 0.3 & 0.00428 & 0.139 & -1.51 & 0.10 & 3830 & 250 & 60.478 & 0.006 & 67.301 & 0.010 & 0.943 & 0.005 & 0.993 & 0.093 \\
\hline
\hline
\end{tabular}
\end{center}
\label{tab02}
\end{table*}
\end{turnpage}

\section{Discussion}
\label{sec:discussion}

Thus far, we have reported measurements of the trajectory of a vortex ring launched at an oblique angle toward a gravity-induced interface separating two fluids.   We have found that the vortex ring experiences a variety of subsequent trajectories after striking the interface.  Perhaps most notably, we have found that, under certain conditions, the vortex ring experiences refraction at the interface.  In this section, we will focus on the interpretation of this phenomenon.  In particular, we will speculate as to whether a generalized law of refraction, similar to Snell's law of refraction for light, might be constructed to describe the refraction of vortex rings.

Recall that one of the early successes of the wave theory of light was its ability to explain Snell's law of refraction by appealing to the variation of the speed of light when crossing a boundary separating media of different densities. Early models of light which emphasized its corpuscular nature proved less convincing in explaining refraction, largely because these models required that light travel more rapidly when entering a denser media, such as water or glass, an ostensibly counterintuitive postulate which was later rejected on experimental grounds~\cite[][]{Smith:1987p2257}. With the advent of Hamilton-Jacobi theory, and later wave mechanics, the refraction of light was eventually reconciled with a corpuscular nature, albeit only in a statistical sense.  Quantum theory, of course, makes no predictions as to the trajectory of an individual photon~\cite[][]{Wolf:1965p2379}.  Nonetheless, recent experiments have revived a significant amount of speculation as to the nature, and even the structure, of the photon~\cite[][]{Nisius:2000p2812, Roychoudhuri:2008yq}.

Regardless of the nature of the photon, it is interesting to inquire whether, apart from a wave, a collective excitation which forms a coherent structure propagating through a medium can exhibit refraction at the boundary of the medium.  Based on our observations, the answer at which we have arrived is: yes, under certain conditions.  

On the one hand, it may seem intuitively clear that a vortex ring would experience refraction when striking a fluid density interface.  After all, a vortex ring is similar to a wavefront in that it is an extended structure, the different spatial regions of which can travel at different velocities.  And spatial variation of wave velocity along a wavefront is precisely the mechanism used to explain refraction of light waves.  

On the other hand, there are a number of problems with such an analogy.  First, unlike with light waves, there is not a unique velocity for a vortex ring in a medium.  That is: the vortex ring slows down over time even in a uniform medium.  This is well known, and is illustrated in our Figs.~\ref{fig04},~\ref{fig06}~and~\ref{fig08}. Second, our results present a bit of a puzzle.  A weak interface can arise either when the Froude number is large or when the Atwood number is small.  A small Atwood number implies a small density difference between the media. If vortex ring refraction is a consequence of a change in vortex ring speed across the interface, then one might expect that a small Atwood number would give rise to a small refraction, since the velocity ratio would be near unity. This, however, is the opposite of what we have observed.  

Now it is still possible that a change in vortex ring speed across the interface can account for the refraction if the buoyancy effects are more capable of masking the larger incipient refraction when the Froude number is small. If this is the case, then perhaps computer simulations of a vortex ring striking a density interface at an oblique angle could lead to a quantitative understanding of the effect of buoyancy on the problem.

Finally, a simplistic application of Snell's law of refraction simply cannot account for all of the data in Tab.~\ref{tab02}.  Recall that column 10 of Tab.~\ref{tab02} depicts the ratio of the sines of the angles of incidence and refraction of the vortex ring.  Pursuing the analogy with the law of refraction, one can inquire whether the ratio of the sines is equal to the ratio of the velocities of the vortex above, and below, the interface.  Column 11 shows $v_i/v_r$, which is the  ratio of the vortex ring ring velocities immediately prior to, and immediately after, impact with the interface.  It appears that  $\sin{\theta_i}/\sin{\theta_r} = v_i / v_r$ holds, to within our experimental uncertainty, for all of the trajectory data except those shown in rows 1, 4, 11 and 14.  This  suggests that a simplistic law of refraction cannot account for all of the data.
 
As an added complication, the vortex ring typically experiences deformation and vorticity shedding at the moment of impact.  Hence, any generalized law of refraction would likely need to account for vortex ring tension, which offers resistance to deformation of its structure. Moreover, we have shown in Sec.~\ref{sec:no_interface} that in some cases the vortex ring trajectory can change abruptly, even in the absence of a fluid density interface, as a result of spontaneous vorticity shedding transverse to the initial vortex ring trajectory.  Indeed it is possible that the observed change in vortex ring trajectory, when it strikes a fluid density interface, might be interpreted as a manifestation of stimulated vorticity shedding.  In any case, regardless of the particular mechanism, the data suggests that, for a given Atwood number, vortex ring refraction occurs at large Froude numbers.

\begin{acknowledgments}
This work was supported by a grant from the Research Infrastructure Program of the NASA Wisconsin Space Grant Consortium.
\end{acknowledgments}


\end{document}